\documentclass[a4paper]{jpconf}
\usepackage{graphicx}
\usepackage{amsmath}
\usepackage{braket}
\usepackage{hyperref}
\usepackage{slashed}
\usepackage{color,soul}
\usepackage{enumitem}
\allowdisplaybreaks
\def\bea#1\eea{\begin{align}#1\end{align}}

\newcommand{\bef}{\begin{figure}[hbt]\centering}
\newcommand{\eef}{\end{figure}}

\newcommand{\f}{\frac}

\def \be  {\begin{equation}}
\def \ee  {\end{equation}}
\def \ba  {\begin{eqnarray}}
\def \ea  {\end{eqnarray}}

\begin{document}
\title{Heavy flavor production in heavy-ion collisions from soft collinear effective theory} 

\author{Zhong-Bo Kang, Felix Ringer, Ivan Vitev}

\address{Theoretical Division, Los Alamos National Laboratory, Los Alamos, NM 87545, USA}

\ead{zkang@lanl.gov, f.ringer@lanl.gov, ivitev@lanl.gov}

\begin{abstract}
We review the approach to calculate open heavy flavor production in heavy-ion collisions based on Soft Collinear Effective Theory (SCET). We include both finite heavy quark masses in the SCET Lagrangian as well as Glauber gluons that describe the interaction of collinear partons with the hot and dense QCD medium. From the new effective field theory, we derive massive in-medium splitting kernels and we propose a new framework for including in-medium interactions consistent with next-to-leading order calculations in QCD. We present numerical results for the suppression of both $D$- and $B$-mesons and compare to results obtained within the traditional approach to parton energy loss. We find good agreement when comparing to existing data from the LHC at $\sqrt{s_{\mathrm{NN}}}=5.02$~TeV and 2.76~TeV.
\end{abstract}

\section{Introduction}

The quark-gluon plasma (QGP) predicted to have existed in the early universe can be reproduced in heavy-ion collisions at RHIC and the LHC. Highly energetic particles and jets that traverse the hot and dense QCD medium provide an ideal probe for this new state of matter. In particular, open heavy flavor production plays a crucial role in probing and understanding the QGP and has received a growing attention by both the experimental and theoretical communities in the past years. The LHC experiments CMS and ALICE have provided high precision data, see e.g.~\cite{CMS:2012vxa,Grelli:2012yv,CMS:2016nrh}, for the nuclear modification factor $R_{AA}$ most commonly used to study the quenching of particle yields in heavy-ion collisions. Interestingly, preliminary results from the CMS collaboration~\cite{CMS:2016nrh} for Pb+Pb collisions at $\sqrt{s_\mathrm{NN}}=5.02$~TeV show that the suppression rate for $D^0$ mesons is the same as for light charged hadrons within the experimental uncertainty.

We present new theoretical calculations beyond the traditional framework of parton energy loss~\cite{Gyulassy:2000fs,Gyulassy:2000er,Djordjevic:2003zk} based on recently developed techniques using Soft Collinear Effective Theory (SCET)~\cite{Bauer:2000ew,Bauer:2000yr,Bauer:2001ct,Bauer:2001yt}. In~\cite{Idilbi:2008vm,Ovanesyan:2011xy}, an effective field theory based on SCET was developed in order to describe highly energetic massless partons traversing the QCD medium. The main idea is to include a Glauber mode that describes the interaction of energetic partons with the QGP. In this work, we extend this calculation by including finite masses for heavy quarks, see~\cite{Kang:2016ofv} for more details. We label the resulting new effective field as SCET$_{\mathrm{M,G}}$. See~\cite{Rothstein:2016bsq} for Glauber modes in a more general context. Following the massless calculations in~\cite{Ovanesyan:2011xy,Ovanesyan:2011kn,Fickinger:2013xwa,Ovanesyan:2015dop}, we derive massive in-medium splitting functions to first order in opacity. With the new in-medium splitting functions, we are able to go beyond the traditional approach to parton energy loss. See also~\cite{Kang:2014xsa,Chien:2015vja,Chien:2015hda,Chien:2016led}. In addition, we introduce a new way for implementing the in-medium corrections consistent with calculations at next-to-leading order (NLO) in QCD~\cite{Jager:2002xm} for inclusive hadron production. This can be achieved by formally defining medium modified fragmentation functions derived from the SCET$_{\mathrm{M,G}}$ splitting functions. As it turns out, the description of the underlying proton-proton baseline plays an important role~\cite{Nason:1989zy,Cacciari:1998it,Kneesch:2007ey,Kniehl:2008zza,Fickinger:2016rfd}.

The remainder of this paper is organized as follows. In section~\ref{sec:two}, we introduce the new formalism to calculate the in-medium corrections consistently at next-to-leading order using the new massive in-medium splitting functions based on SCET$_{\mathrm{M,G}}$. In section~\ref{sec:three}, we present numerical results for the suppression rates for both $D$- and $B$-mesons and compare to data from the LHC. In section~\ref{sec:four}, we conclude and give an outlook.

\section{Massive in-medium splitting functions and their application to $\mathrm{PbPb}\to HX$ \label{sec:two}}

\begin {figure*}[t]
\begin{center}
\vspace*{10mm}
\includegraphics[width=0.4\textwidth,trim=1cm 2cm 1cm 1cm ]{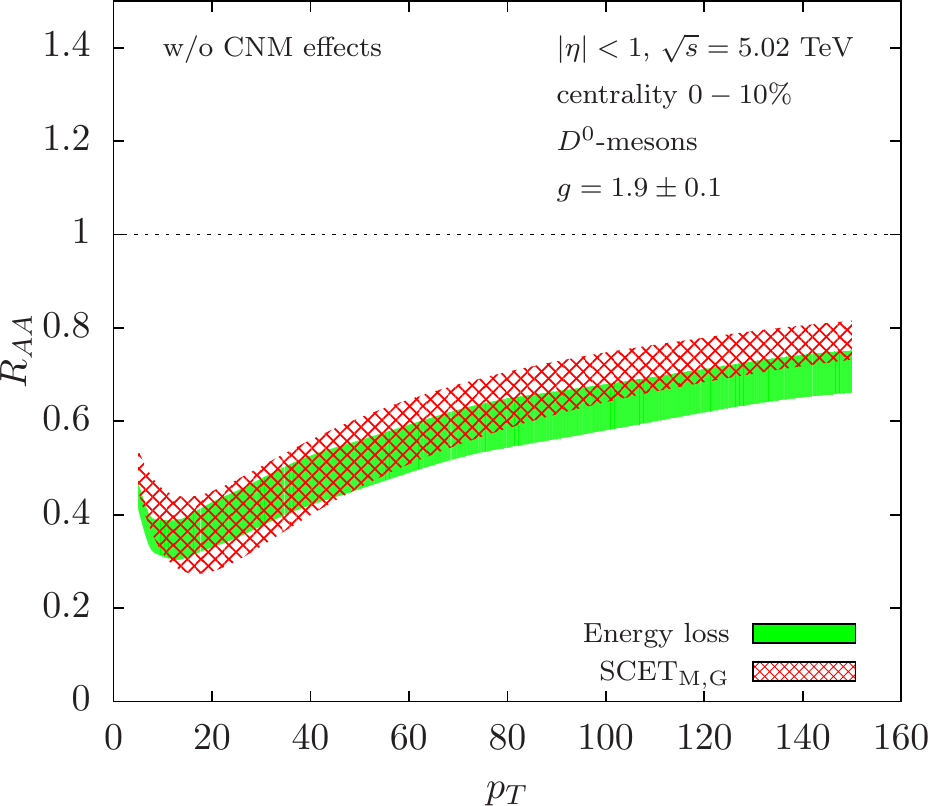} 
\hspace*{2cm}
\includegraphics[width=0.4\textwidth,trim=1cm 2cm 1cm 1cm ]{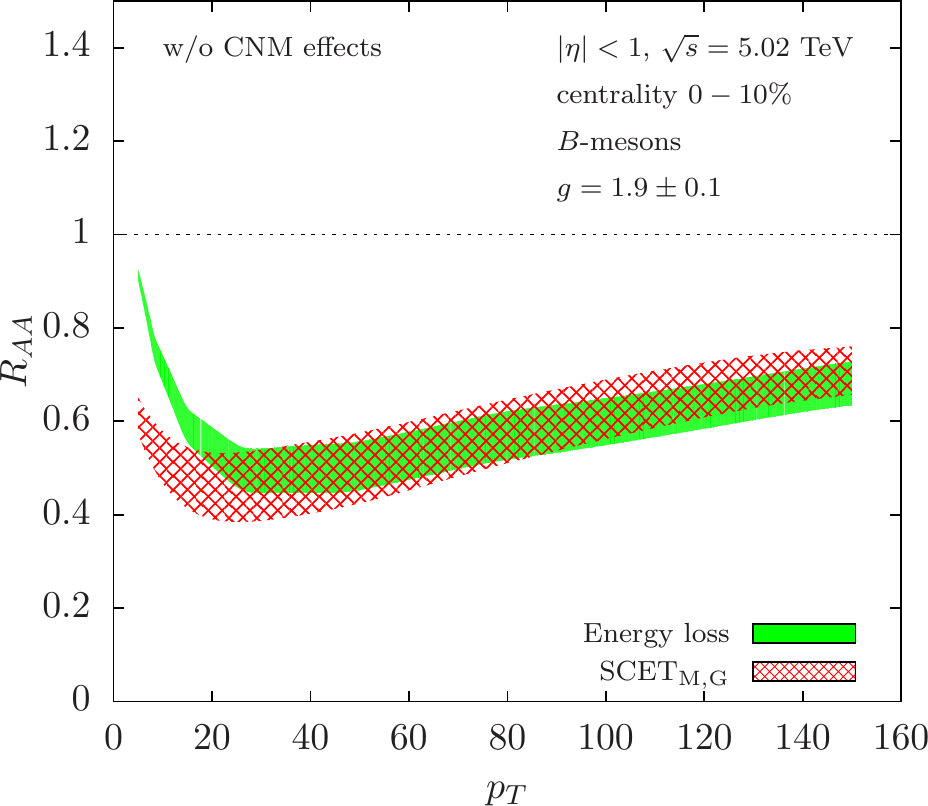} 
\end{center}
\vspace*{1.cm}
\caption{\label{fig:RAA-A} Suppression of $D^0$-mesons (left) and $B$-mesons (right) in central PbPb collisions at $\sqrt{s_{\mathrm{NN}}}=5.02$~TeV. We show results obtained within the traditional approach to parton energy loss (green) and the new results using SCET$_{\mathrm{M,G}}$ (red).}
\end{figure*}

The SCET$_{\mathrm{M,G}}$ Lagrangian is given by a sum of the massive SCET Lagrangian in the vacuum~\cite{Leibovich:2003jd} and the Glauber gluon interaction terms as in the massless case~\cite{Ovanesyan:2011xy}. The medium interaction terms are not modified due to the finite mass effects, which can be obtained from power counting. See~\cite{Kang:2016ofv} for more details. From the resulting Lagrangian, we can directly derive the massive vacuum splitting functions $Q\to Qg$, $Q\to gQ$ and $g\to Q\bar Q$ which were first obtained using traditional perturbative methods in QCD~\cite{Catani:2000ef}. The in-medium results are obtained by taking into account single- and double-Born diagrams and by summing over the number of scattering centers. As it turns out, the in-medium splitting functions have a similar structure as the massless results in~\cite{Ovanesyan:2011xy,Ovanesyan:2011kn}. Since the final expressions are quite lengthy, we refer the interested reader to~\cite{Kang:2016ofv}. In the soft emission limit, it is possible to reproduce the splitting functions obtained in traditional energy loss calculations for heavy quarks~\cite{Djordjevic:2003zk}.

The interactions with the QGP affect only the partons after the hard-scattering event. Following the methods developed in~\cite{Collins:1988wj}, we consider both real and virtual corrections. The relevant part in the cross section for a fragmenting quark or gluon $i$ can be schematically written as
\be\label{eq:medNLO}
\sum_j\sigma^{(0)}_{i} \otimes {\cal P}_{i\to jk} \otimes D_j^{H} \, ,
\ee
where $\sigma^{(0)}_i$ is the leading-order production cross section for quarks or gluons. The ${\cal P}_{i\to jk}$ describe the splitting process $i\to jk$ and are given by the sum of vacuum and in-medium splitting functions
\be\label{eq:vac+med}
{\cal P}_{i\to jk}(z,\mu)={\cal P}^{\mathrm{vac}}_{i\to jk}(z,\mu)+{\cal P}^{\mathrm{med}}_{i\to jk}(z,\mu) \, .
\ee
The first terms in Eqs.~(\ref{eq:medNLO}) and (\ref{eq:vac+med}) are simply part of the NLO calculation in the vacuum~\cite{Jager:2002xm}. The second term in Eq.~(\ref{eq:vac+med}) gives the fixed order in-medium correction and can be formally considered as a medium modified fragmentation function. See~\cite{Kang:2016ofv} for a more detailed derivation.

\section{Numerical results and comparison to data \label{sec:three}}

\begin {figure*}[t]
\begin{center}
\vspace*{10mm}
\includegraphics[width=0.4\textwidth,trim=1cm 2cm 1cm 1cm ]{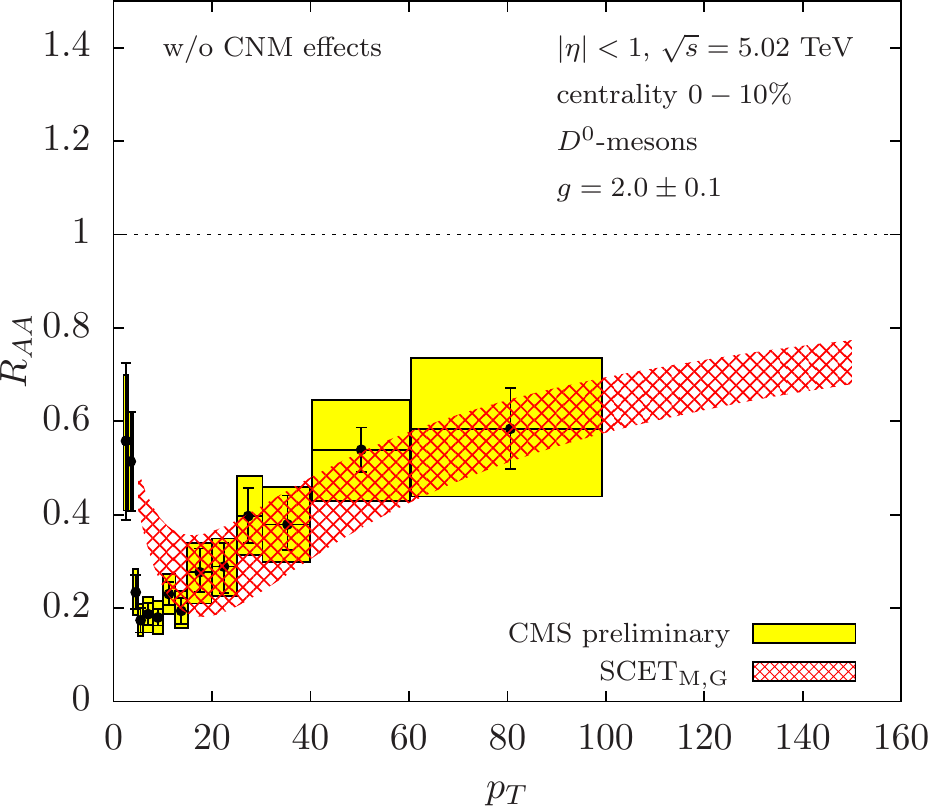} 
\hspace*{2cm}
\includegraphics[width=0.4\textwidth,trim=1cm 2cm 1cm 1cm ]{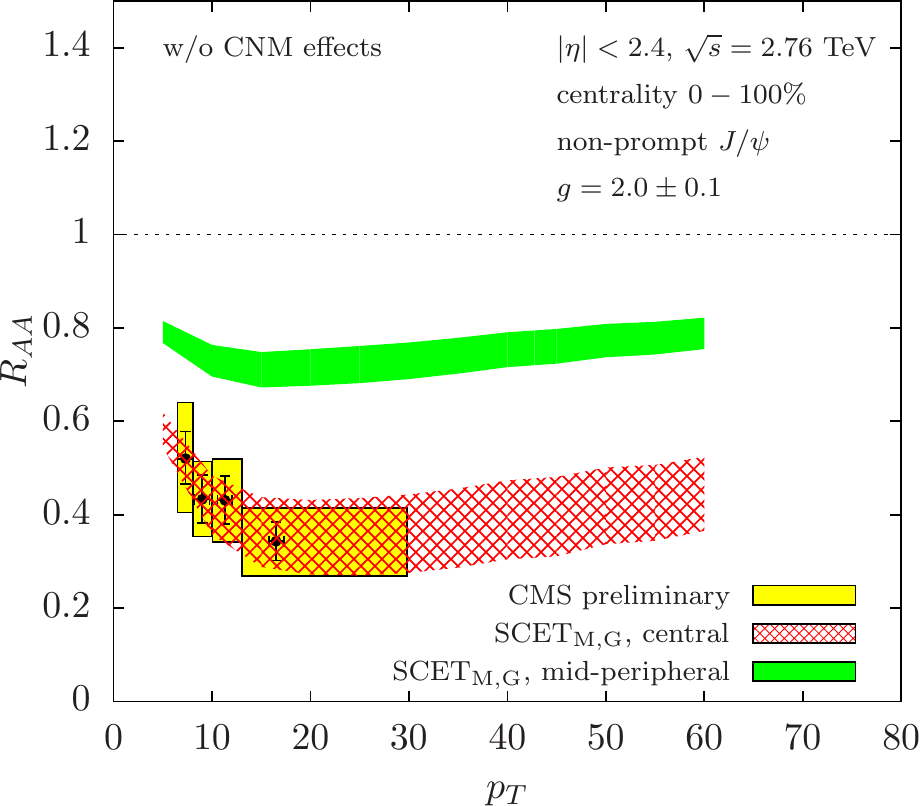} 
\end{center}
\vspace*{1.cm}
\caption{\label{fig:RAA-B} Comparison of SCET$_{\mathrm{M,G}}$ based results with preliminary CMS data for $D^0$-mesons~\cite{CMS:2016nrh} (left) and non-prompt $J/\psi$ which originate from $B$-mesons~\cite{CMS:2012vxa} (right).}
\end{figure*}

In this section, we present results for the nuclear modification factor $R_{AA}$ which is defined as the ratio of the PbPb and the proton-proton cross sections
\be
R_{AA}=\f{d\sigma^H_{\mathrm{PbPb}}/d\eta dp_T}{\braket{N_{\mathrm{coll}}} d\sigma^H_{pp}/d\eta dp_T} \,  ,
\ee
where $\braket{N_{\mathrm{coll}}}$ is the average number of binary nucleon-nucleon collisions for a given centrality. Note that for the proton-proton baseline, we choose to use the Zero Mass Variable Number Scheme (ZMVFNS) at NLO and the fragmentation functions of~\cite{Kneesch:2007ey,Kniehl:2008zza}. We find that this scheme can be applied even for relatively low $p_T$ of the observed hadron $H$. For our numerical results, we do not take into account Cold Nuclear Matter (CNM) effects. Firstly, we present a comparison to results obtained within the traditional approach to parton energy loss in Fig.~\ref{fig:RAA-A}. Secondly, in Fig.~\ref{fig:RAA-B} we compare our SCET$_{\mathrm{M,G}}$ results to preliminary CMS data for $D^0$ mesons~\cite{CMS:2012vxa} (left) at $\sqrt{s_{\mathrm{NN}}}=5.02$~TeV and for non-prompt $J/\psi$ originating from $B$-hadrons~\cite{CMS:2016nrh} (right) at $\sqrt{s_{\mathrm{NN}}}=2.76$ TeV. Note that the non-prompt $J/\psi$ data is in fact only for minimum bias events instead of fixed centrality. Therefore, we present our calculation for both central as well as mid-peripheral collisions. We find that our  central result (0-10\% centrality) agrees very well with the data. The minimum bias results are dominated by central collisions as they are weighted with $\braket{N_{\mathrm{coll}}}$. We note that the heavy-ion results crucially depend on whether the heavy meson is produced by a fragmenting heavy quark or a gluon. In general, gluons 
 more energy than (heavy) quarks when undergoing multiple scatterings in the medium before they eventually fragment into the observed heavy meson. See~\cite{Kang:2016ofv} for more details.

\section{Conclusions \label{sec:four}}

We derived a new version of Soft Collinear Effective Theory including both finite heavy quark masses and the interaction of collinear partons with the medium that are mediated by Glauber gluon exchange~\cite{Kang:2016ofv}. From the new effective field theory, we are able to deduce the in-medium massive splitting functions for $Q\to Qg$, $Q\to gQ$ and $g\to Q\bar Q$ to first order in opacity. In addition, we introduced a new formalism to include in-medium effects consistent with calculations at next-to-leading order in QCD. We presented numerical results for the suppression of $D$- and $B$-mesons in heavy-ion collisions. We found good agreement in comparison to existing data from the LHC at $\sqrt{s_\mathrm{NN}}=5.02$~TeV and 2.76~TeV. In the future, we plan to extend the current framework of the in-medium modification to inclusive jet production~\cite{Kang:2016mcy}.
 
\section*{Acknowledgments}

We would like to thank Grigory Ovanesyan for his collaboration at the early stages of this work. In addition, we would like to thank Yen-Jie Lee, Emanuele Mereghetti, Rishi Sharma, Hongxi Xing and Zhiqing Zhang for helpful discussions. This work is supported by the U.S. Department of Energy under Contract No. DE-AC52-06NA25396, in part by the LDRD program at Los Alamos National Laboratory.

\end{document}